\documentclass[a4paper,12pt]{article}
\usepackage{epsf}

\newcommand{\rand}{\qquad\mathrm{and}\qquad}
\newcommand{\WDW}{Wheeler--DeWitt}
\newcommand{\KS}{Kantowski--Sachs}
\newcommand{\reell}{\mbox{I$\!$R}}

\begin{document}
\sloppy

\vspace*{-13mm}\hfill PITHA 95/36
\newline \vspace*{1mm}\hspace*{10cm}\hfill April 1996
\newline \vspace*{1mm}\hspace*{10cm}\hfill gr-qc/9604036

\vspace*{15mm}
\begin{center}{\bfseries\huge Tunneling of Macroscopic Universes}
\end{center}
\vskip 1.0cm
\begin{center}
{\large Heinz--Dieter Conradi \\
Institute for Theoretical Physics E, RWTH Aachen \\
Sommerfeldstr. 26-28, 52056 Aachen, Germany \\
Email: conradi@physik.rwth-aachen.de
}
\end{center}

\vspace*{5mm}
\begin{center}
\end{center}
\vspace*{5mm}

\begin{abstract}
  The meaning of `tunneling' in a timeless theory such as quantum
  cosmology is discussed. A recent suggestion of `tunneling' of the
  macroscopic universe at the classical turning point is analyzed in
  an anisotropic and inhomogeneous toy model. This `inhomogeneous
  tunneling' is a local process which cannot be interpreted as a
  tunneling of the universe.


\end{abstract}

\thispagestyle{empty}

\noindent{\Large\bfseries 1. Timelessness and tunneling}
\typeout{Timelessness and tunneling}

\noindent
Quantum gravity and quantum cosmology have the reputation of not
describing any accessible experiments and for not having any
consequences after the Planck era. However, it has been suggested that
quantum cosmology could be responsible for some phenomena of everyday
physics as the arrow of time and the classical appearance of our
universe \cite{zeh_book,barbour_94a}, the smallness of the
cosmological constant \cite{coleman_88,hawking_90a}, or the onset of
inflation \cite{fischler_etal_90b,farhi_etal_90}.

As a prominent subject over the years, tunneling processes have been
considered for different purposes in quantum cosmology. In a recent
paper D\c{a}browski and Larsen discussed the tunneling of a
recollapsing universe at the classical turning point into an expanding
state with increased scale factor \cite{dabrowski_larsen_95a}. Similar
models have been studied e.g. also by \cite{rubakov_84} and, for the
early stage of the universe, by
\cite{fischler_etal_90b,farhi_etal_90}. In this letter I will comment
on the meaning of `tunneling' in quantum cosmology and the model of
\cite{dabrowski_larsen_95a} is substantially generalized.

The \WDW\ equation $H\Psi=0$ which governs quantum cosmology may be
regarded as analogous to the stationary Schr\"odinger equation. The
similarity is particularly striking for the popular one dimensional
models in which the universe is described by the scale factor
$a$ only. The universe then resembles a particle in one dimension
and the \WDW\ equation reads
\begin{equation}
  \left[-\partial_a^2 + V(a)\right]\Psi\ =\ 0 \ ,\label{one}
\end{equation}
with $a\in\reell_+$. However, this similarity is misleading since,
among other differences, there exists an external time parameter in
ordinary quantum mechanics in contrast to quantum cosmology.

The quantum mechanical time parameter is crucial even for the most
simple examples of tunneling. Consider the reflection and tunneling of
a wave function at a square potential barrier. This situation is
described by superposing an ingoing wave with a reflected wave in one
free region while in the free region on the other side of the
potential barrier the wave function consist solely of an {\itshape
  outgoing\/} wave, the amplitude of which determines the tunneling
probability. The concepts of in-- or outgoing waves are justified by
either considering a wave packet by superposing different energy
eigenstates the sum of which resembles a particle moving {\itshape in
  time}. (Due to time--reparametrization invariance there is only the
`zero energy solution' to deal with in quantum cosmology. In one
dimension it is thus impossible to form wave packets.) Or more simply
by observing that the crests of the solitary waves are {\itshape
  moving\/} with $\pm kx-\omega t=2\pi n$. The necessity of a time
parameter is, however, clear from the onset since the very notion of
tunneling assumes a state changing in time: While initially there is
no particle in some region (and classically there will never be one),
there is a finite probability to find one subsequently.

It is worthwhile to point out that a tunneling {\itshape
  probability\/} means probability of a tunneling `event' of a
particle and thus implies a measurement. The concept of tunneling thus
presupposes both a time parameter and a theory of measurement.
Concerning the first point, it does not help that there is a conserved
Klein--Gordon current in quantum cosmology in more than one dimension,
since its sign cannot be fixed in the absence of an external time.
Denoting something as outgoing as e.g. in the definition of the
`tunneling wave function' \cite{vilenkin_82} would completely
arbitrary fix a direction on the configuration space.

In the case of a high and broad potential barrier the situation 
simplifies, since the wave function in the forbidden region can be 
approximated by the exponentially suppressed solution only. In
quantum mechanics the ratio of the (squared) wave function at the 
beginning and at the end of the forbidden region gives the tunneling 
probability. The same procedure is usually adopted in quantum 
cosmology as e.g. in \cite{fischler_etal_90b} in order to calculate 
the tunneling probability of a bubble of false vacuum between two 
classically allowed regions in a `free lunch' process. Accepting for 
the moment this interpretation, one arrives at similar conclusions in 
the semiclassical limit of quantum cosmology as in ordinary quantum 
mechanics.

Usually, the situation is more complicated than in the examples above,
as e.g. in the alpha decay when one of the two classically allowed
regions is restricted to a finite region in space. But while in that
case it is still possible to have a purely outgoing wave outside the
nucleus, this breaks down if both classically allowed regions are
finite. (D\c{a}browski and Larsen used this kind of potential for
analyzing the tunneling probability at the classical turning point of
a FRW universe with several matter sources.) In Fig.~1 the first
situation is depicted by the dashed line while the case with two
bounded regions is shown by the solid line.
\parbox{5mm}{}
\begin{figure}[hbt]
\parbox{90mm}{\epsfxsize=85mm
  \epsffile{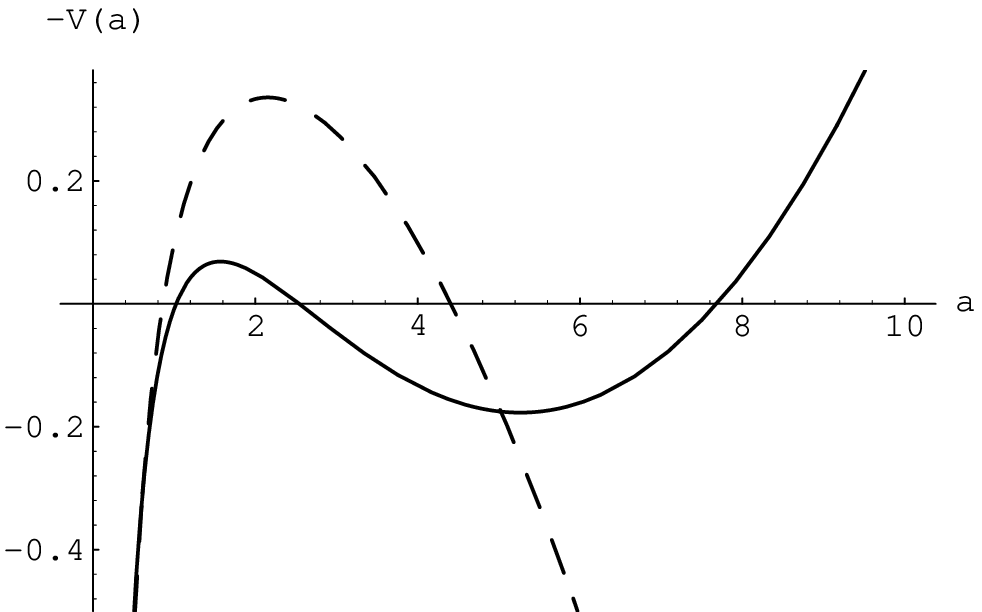} } \hfill\parbox{70mm}{
\label{bild} \footnotesize
  \ \newline
  Fig.~1:
  The solid line shows the kind of potential used by D\c{a}browski and
  Larsen. The second classically allowed region is due to domain walls
  and a negative cosmological constant. In order to get a `tunneling
  universe', obviously, the ascending part of the potential is not
  necessary. In Sec.~2 a potential is used which is depicted by the
  dashed line. The decreasing is due to a positive cosmological
  constant.
}\hfill\parbox{5mm}{}
\end{figure}
In order to calculate the tunneling probability it is thus not
sufficient to consider stationary waves only. The calculation is much
more involved as can be seen e.g. by the deviation from the
exponential decay law for the nuclear decay in a box
\cite{haroch_kleppner_89}. Without the aid of the external time
parameter the situation in quantum cosmology is quite unclear. One
cannot even use the above mentioned comparison of the wave function at
both ends of the potential barrier as a formal tool, since there will
be no purely exponentially suppressed solutions.

In view of the aforementioned problems I will use the notion 
`tunneling' only as a formal notion {\itshape provided\/} there exists 
a purely exponentially decaying solution: The ratio of the squared 
wave function at the beginning and at the end of the exponential 
region is defined as `tunneling probability'. (Note that the
exponentially increasing wave function will also be a solution ---
`reversed tunneling' --- and arbitrary superpositions of this two 
basic solutions.)

This formal concept may only be regarded as a corresponding to a `real
process' if there is (at least) a time parameter and a theory of
measurement in quantum cosmology. In more complicated examples as e.g.
the one of Fig.~1 one furthermore has to be rather careful with the
calculation of the probabilities. One may try to circumvent
some of these difficulties by introducing a semiclassical time
parameter e.g. due to a decoherence process or due to a
Born--Oppenheimer type of approximation. Usually this point of view is
put forward by mentioning that the tunneling occurs in a region where
the semiclassical approximation should hold, although Kiefer and Zeh
have argued that this argument is not sufficient \cite{kiefer_zeh_95}.
If, however, decoherence could be used to define a semiclassical time,
it is expected to simultaneously suppress the tunneling process by a
Zeno--type effect \cite{caldeira_leggett_83b}. I will come back to
these issues in Sec.~3 when the results of the calculations in the toy
model are discussed. As frequently stressed above, the notion of
tunneling makes sense only if there is a time parameter. Since this is
certainly not the case in the Planck era one cannot sensibly speak of
`tunneling from nothing' of the universe which is frequently used as a
quantum alternative to the big bang.

The preceding considerations implicitly assume the so called
na$\ddot{\i}$ve interpretation of the wave function which itself
relies on the similarity of the \WDW\ equation with the
Schr\"odinger equation. That is, I assume the wave functions to
be normalizable on the {\itshape whole\/} (unconstrained)
configuration space or that at least a conditional probability
can be used. Any other interpretation in quantum cosmology which
works on the reduced phase space or isolates a time parameter is
meaningless in the example of the FRW universe with
phenomenological matter. This is because in these interpretations
there is but one physical state. But even for more realistic
examples in which the time parameter is given by some function of
the volume, it does not make sense to compare volumes. As already
mentioned above, even if a physical time has been introduced one
has to establish a theory of measurement in quantum cosmology
before one could sensibly speak of tunneling.

\ 

\noindent{\Large\bfseries 2. An anisotropic and inhomogeneous model}
\typeout{An anisotropic and inhomogeneous model}

\noindent

The \KS\ model as the most simple anisotropic and even
inhomogeneous model has the advantage that it can be solved
exactly. The homogeneous version of the model combines spherical
symmetry with a translational symmetry in the `radial' direction.
(First the homogeneous model is discusses before the
translational symmetry is relaxed, see below.) The spacelike
hypersurfaces of constant times are therefore cylinders
\begin{equation}
  ds^2\ =\ z^2\,dr^2 + b^2\,d\Omega^2 \ .\label{metric}
\end{equation}
Here $b(t),z(t)\in\reell_+$; $b(t)$ is the surface measure of the
two--spheres with metric $d\Omega^2$ and $z(t)$ measures the
spacelike distance between them; $r$ is the radial coordinate.

The model with positive cosmological constant $\Lambda$ and
pressureless dust is considered here. It is well known that it is
only the presence of matter which renders the disklike
singularity ($z\to0$) from a mere coordinate singularity which
indicates the incompleteness of the model into a curvature
singularity. The dust is described by the parameter $z_{\mathrm
  m}=\rho zb^2=const$ (analogous to $a_{\mathrm m}=\rho a^3$ in
the FRW model). For homogeneous models this approach is
essentially equivalent to a more sophisticated one which starts
from a Lagrangian for the dust degrees of freedom, see e.g.
\cite{brown_kuchar_95a} and the literature cited therein.
Although dust as a matter source in quantum cosmology is
unsatisfactory, it does here lead to a toy model which is
calculable and complete. In addition, the present context of a
macroscopic universe justifies this phenomenological description.

The following account of the classical dynamics of this well
known model is rather cursory. More details can be found e.g. in
\cite{conradi_95a,laflamme_shellard_87} and the literature cited
therein. That time parameter is used which is equivalent to
conformal time in the FRW model ($N=b$). The dynamics is
determined by the Hamiltonian constraint 
\begin{equation}
  z\dot{b}^2\ +\ 2\dot{z}b\dot{b}\ +\ b^2 \left( z\ -\ z_{\mathrm m}
  - \Lambda zb^2 \right)\ =\ 0
\ ,\label{hamilton}
\end{equation}
(written in the configuration variables and their velocities) and by
one of the equations of motion
\begin{equation}
  \dot{b}^2 + b - b_{\mathrm m} - \frac 13 \Lambda b^3 \ =:\
  \dot b^2 - \tilde V_{b_{\mathrm m}}(b)\ =\ 0
\ .\label{eqb}
\end{equation}
Here the equation of motion for $b(t)$ has already been integrated,
with $b_{\mathrm m}$ as an {\itshape arbitrary\/} constant of motion.
The equation of motion for the scale factor of the closed FRW model is
identical to Eq.~(\ref{eqb}), $\ddot a(t)=\ddot b(t)$, but the
corresponding constant $a_{\mathrm m}$ is fixed by the matter content.
In the \KS\ model the matter content instead fixes the constant of
integration of $z(t)$ as indicated by the notation.

The classically forbidden regions in this model are determined by
$\tilde V_{b_{\mathrm m}}(b)\le0$. Note that there is no
forbidden region on configuration space due to the Hamiltonian
(\ref{hamilton}) since its kinetic term is indefinite.
Consequently, the above defined forbidden region is defined for
each classical solution separately, because each solution is
uniquely represented by the `effective mass' $b_{\mathrm m}$.
There are two classically allowed regions only if the
cosmological constant is positive and smaller than the Einstein
value: $0<\Lambda<4/(9b_{\mathrm m}^2)$. This potential which is
depicted by the dashed line in Fig.~1 is similar to the situation
in \cite{dabrowski_larsen_95a}, see the solid line in the same
figure, except that it is not increasing for large values of $b$.

The one--dimensionality of Eq.~(\ref{eqb}) suggests, furthermore,
the possibility of an inhomogeneous $z(t,r)$. Geometrically, this
looks reasonable because a cylinder is defined only by the
homogeneity of the surface measure $b$ while an inhomogeneous
spacelike distance $z$ between the two--spheres would not deform
it. It turns out that an inhomogeneous $z(t,r)$ is even
dynamically consistent for pressureless dust as a matter content,
as was first noticed by Ellis \cite{ellis_67}. A rigorous proof
is possible by using the equations of motion for the general
spherically symmetric model as is performed in the appendix.
There it follows directly, since $z'$ is always tied to $b'$ and
thus does not appear in the \KS\ model. Due to this absence of
any spatial derivatives, states at different points do not
interact: The Hamiltonian (\ref{hamilton}) remains thus
essentially unchanged 
\begin{equation}
  z(t,r)\dot{b}^2(t)\ +\ 2\dot{z}(t,r)b(t)\dot{b}(t)\ +\ b^2(t) 
  \left( z(t,r)\ - z_{\mathrm m}(r) - \Lambda z(t,r)b^2(t) 
  \right)\ =\ 0
\ ,\label{r-hamilton}
\end{equation}
One thus has a one parameter set of homogeneous solutions
$b=b(t)$ which then determines uniquely the solution $z=z(t,r)$
the inhomogeneity of which is due to the inhomogeneity of the
dust.

In the recollapsing region one gets a qualitatively correct
picture of the dynamics by considering the exact solutions for
$\Lambda=0$:
\begin{equation}
  b(t) = b_{\mathrm m} \sin^2\left(\frac{t}{2}\right)
  \rand 
  z(t,r) = K(r)\cot\left(\frac{t}{2}\right) +
  z_{\mathrm m}(r)\left( 1 - 
  \frac{t}{2} \cot\left(\frac{t}{2}\right) \right)
\ ,\label{dynamics}
\end{equation}
with $t\in[0,2\pi]$ and where $K$ is a physically meaningless
`constant' of time--integration (different functions of $K$ are
identified by a redefinition of the radial coordinate and by
simultaneously changing the dust potential). The other
classically allowed region for large $b$ can be approximated by
considering only the cosmological term in the potential.
Obviously, the inhomogeneity of $z(t,r)$ is due to inhomogeneous
dust $z_{\mathrm m}(r)$.

As usual Dirac's quantization scheme is used by turning the variables
$b$, $z$ and their conjugate momenta into operators which satisfy the
standard commutation relations. The Hamiltonian constraint
(\ref{hamilton}) is turned into an operator which annihilates the
physical states: $\hat H\Psi=0$, the so called \WDW\ equation. In the
configuration space representation it reads
\begin{equation}
  -2zb^2\hat{H}\Psi= \left[ 
  z^2\partial^2_z + kz\partial_z - 2zb\partial_b\partial_z 
  + z^2b^2 \left( 1-\Lambda b^2 \right) - zz_{\mathrm m}b^2
  \right] \Psi(b,z) = 0
\ ,\label{wdw} 
\end{equation}
where $\partial_z$ and $\partial_b$ are the partial derivatives
with respect to $z$ and $b$, respectively. Factor ordering is
partially left open as indicated by the parameter $k$. The
Laplace--Beltrami ordering is given by $k=1$. This equation can
exactly be solved for a more general potential which contains
arbitrary functions of $b$ multiplied by $z$ and $z^2$
\cite{conradi_95a}.

Although $z(r)$ is a field it suffices to consider the
minisuperspace \WDW\ equation (\ref{wdw}) at each point $r$ since
there are no partial derivatives with respect to $r$ (and because
of the simple structure of the solutions).  Quantum fluctuation
are expected to result in interaction between different points.
But this effect can only be considered in the context of a more
general model as is common to this kind of problems.

In order to solve the \WDW\ equation it is convenient to
introduce the operator $\hat b_{\mathrm m}:=-\frac
1b\partial_z{}^2 +(b-\Lambda b^3/3)$ since $[\hat b_{\mathrm
  m},\hat H]=0$.  As indicated by the notation, the eigenvalues
of this operator is the effective mass for $b$. The eigenvalue
equation $\hat b_{\mathrm m}\Psi_{b_{\mathrm m}} = b_{\mathrm
  m}\Psi_{b_{\mathrm m}}$ can easily be solved, and one finally
ends up with the following set of exact solutions for the \WDW\
equation \cite{conradi_95a}, 
\begin{equation} 
  \Psi_{b_{\mathrm m}}(z,b) = 
  \sqrt{\frac{b^{k-1}}{|\tilde V_{b_{\mathrm m}}(b)| }} 
  \exp\left\{\pm\mathrm i\left(
  z\sqrt{ b\tilde V_{b_{\mathrm m}}(b)} + \frac{z_{\mathrm m}}{2}
  \int^b\sqrt{\frac{b'}{\tilde V_{b_{\mathrm m}}(b')}}\ \mathrm db'
  \right)\right\}
\ .\label{psiks} 
\end{equation}
Alternatively, one can consider the solutions which are given by
a superposition of the exponential with plus and minus sign to
form e.g.  the `$\cos$' and the `$\sin$' (the `$\cosh$' and
`$\sinh$' in the forbidden region). The integral in the
exponential can be expressed in terms of elementary function for
$b_{\mathrm m}=0$ (that is for large values of $b$) and for
vanishing $\Lambda$ (that is for small values of $b$).  Neither
case is appropriate here.

In order to analyze the formal tunnel probability one has to
calculate the ratio $prob:=|\Psi(b_1)|^2/|\Psi(b_2)|^2$ where
$b_1$ and $b_2$ denote the beginning and the end of the
classically forbidden region, respectively. While, the prefactor
of the wave function is divergent at the borderline of the
classically allowed region it cancels in $prob$, since
$\tilde{V}_{b_{\mathrm m}}(b)$ possesses three distinct, simple
zeros (the third one is negative and has no physical
significance). The first term in the exponent vanishes for
$b=b_1,b_2$. There remains the second term in the exponential
which is a definite integral between $b_1$ and $b_2$. This
integral remains finite since at both limits of integration the
integrand is approximately $1/x$ and the infinities at the two
boundaries cancel each other.

If one is interested in a `tunneling' from the recollapsing
region into the region right from the potential barrier, one has
to choose the minus sign in the exponent of the wave function.
(Note that the Hartle--Hawking wave function has the opposite
sign \cite{laflamme_shellard_87} and describes thus a `tunneling'
{\itshape into\/} the recollapsing region.) The final result then
is given by: $prob = (b_1/b_2)^{(k-1)} \exp\{- z_{\mathrm m}
F(b_{\mathrm m},\Lambda)\}$, where the number $F(b_{\mathrm
  m},\Lambda)$ is the definite integral. Inserting some simple
values: $b_{\mathrm m}=1$, $\Lambda=3/8$ gives $b_1=\sqrt 5-1$
and $b_2=2$; the evaluation of the integral gives: $F(b_{\mathrm
  m},\Lambda)\approx 5.106$. Choosing furthermore the
Laplace--Beltrami factor ordering, $k=1$, one finds
\begin{equation}
  prob\ \approx\ \exp\{ -5.106 z_{\mathrm m}(r)\} \ .
\label{prob}
\end{equation}
The simple structure of this result relies on the semiclassical
structure of the exact wave function plus the canceling of the
prefactor in $prob$.

\ 

\noindent{\Large\bfseries 3. Results and remarks}
\typeout{Results and remarks}

\noindent 
A remarkable feature of the above `tunneling probability' is the
way it depends on the matter content: The logarithm of $prob$
depends linearly on $z_{\mathrm m}$. Apart from factor ordering
effects the tunneling probability equals one for the vacuum
model\footnote{ In this way the model automatically takes into
  account that the vacuum model does not possess forbidden
  regions: It describes the dynamical regions of a black hole (in
  de~Sitter space).  }.  Since the matter does {\itshape not\/}
influence the tunneling barrier one has the analogue of a
particle with mass $z_{\mathrm m}$ (squared) running against a
potential barrier.  However, different points $r$ in the \KS\ 
model will `tunnel' individually since they do not interact. This
situation is analogues to a cloud of non--interacting particles.
In more realistic models one might think of weakly coupled points
(perhaps galaxies), which nevertheless will behave mainly
independent in the `tunneling process'. This {\itshape local\/}
process clearly cannot be interpreted as a tunneling of a whole
universe.

How is this process to be interpreted? Due to the timeless nature
of quantum cosmology, and in particular since a semiclassical
time might not be defined at the turning point, both `universes'
(that is both classically allowed regions) exist
`simultaneously'. In this case one might interpret the tunneling
as a quantum wormhole\footnote{ A quantum wormhole is formally
  defined by an Euclidean region between two separated classical
  regions.  }.  However, the same argument of timelessness tells
one, that there is no classical observer at the turning point, no
incoming and no outgoing wave. The world is completely quantum
and consequently one cannot speak of a tunneling process or of
wormholes. If, on the contrary, the tunneling occurs as a change
`in time', e.g. defined by decoherence, the variable $b$ in some
space regions changes from $b_{\mathrm{max}}$ of the recollapsing
solution to $b_{\mathrm{min}}$ of the solution right from the
potential barrier, with $b_{\mathrm{min}} > b_{\mathrm{max}}$.
This is similar to the `free lunch' process in the very early
universe \cite{fischler_etal_90b,farhi_etal_90}. These tunneling
regions behave dynamically different from their environment since
they are now expanding instead of recollapsing and because of the
sudden change of the $b$-variable. This results in a destruction
of the cylinder geometry of the universe.

It has been emphasized by D\c{a}browski and Larsen that this
tunneling process does not require any change in the matter
content as e.g. a change of the vacuum of the involved fields.
However, Rubakov has suggested that the matter content may be
changed due to the very tunneling process \cite{rubakov_84}. He
considered a scalar field, conformally coupled to the FRW model,
and observed that the tunneling probability is enhanced with
growing particle content. Probably, this remark will apply too
when only small parts of the universe are involved. However, the
calculation for the \KS\ model is much more involved than that
for the FRW model.

A comparison of $prob$ for the \KS\ and for the FRW model shows
another difference than that of inhomogeneity. Starting from the
\WDW\ equation for the FRW model $[\partial_a^2 + \frac 13\Lambda
a^6 + a^4 - a_{\mathrm m}a^3]\Psi=0$ one gets (in WKB
approximation) 
\begin{equation}
  prob_{_{\mathrm{FRW}}} = \exp\left\{ 
  -2\int^{a_2}_{a_1}a^3\tilde{V}(a,a_{\mathrm m},\Lambda)\mathrm da
  \right\}
\ ,
\end{equation}
where $\tilde{V}$ is defined as in the \KS\ model. This result is
quite different from the expression (\ref{prob}) for the \KS\ 
model, since in the FRW model $a_{\mathrm m}$ is determined by
the matter content. The analogue for the tunneling in the FRW
model is that of a particle running against a barrier the
{\itshape width\/} of which is fixed by the matter content.

It was shown in \cite{conradi_95a} that one can insert a \KS\ 
cylinder between two FRW half--spheres. The `tunneling
probability' in this compact model changes drastically at the
borderline between the different parts. However, since the dust
content in the FRW model fixes the constant of integration
$b_{\mathrm m}$ of the \KS\ region, the tunneling probability in
the \KS\ region is a function of the FRW dust, too. The
functional dependence is even similar in both regions.

It has been argued in the last section that the calculation of
$prob$ is not spoiled by the divergences of the wave function
(\ref{psiks}).  One can even get rid of these divergences by
considering superpositions. (In contrast to the delta--functional
like solutions (\ref{psiks}) it is e.g. possible to consider
`free waves', the form of which reveals nothing of classically
forbidden regions \cite{conradi_etal_96}.) This is to be done in
any case since usually a universe is not represented by one
eigenfunction but by a wave packet. The above calculation of
$prob$ is nonetheless necessary since the forbidden regions are
defined for fixed values of $b_{\mathrm m}$ only. With other
words, every single component of the superposition `tunnels'
independently. Moreover, in more complicated models there might
not be a sharply peaked wave packet at the classical turning
point due to interference effects between the `incoming' and the
`outgoing' part \cite{kiefer_88b}. This is in agreement with the
point of view that there is no tunneling because the wave
function at this point is completely quantum.

There might be thus two effects of quantum cosmology at the
classical turning point of the universe: The breakdown of the
semiclassical approximation and a local change into an expanding
state. One might try to circumvent the breakdown of the
semiclassical approximation by considering decoherence. If this
worked (Kiefer and Zeh have expressed their doubts that it does
\cite{kiefer_zeh_95}), the increased classicallity of the
universe would further suppress the tunneling probability
\cite{caldeira_leggett_83b}. Both quantum effects would then be
lost simultaneously, by the same token.

\

\noindent{\Large\bfseries Acknowledgment}
\typeout{Acknowledgment}

\noindent
I thank Dieter Zeh and Claus Kiefer for their enlightening remarks,
Hans Kastrup for his criticism, and Henning Wissowski and Thomas
Strobl for supporting me in the difficult task of making this text
intelligible. This work was supported by the DFG.

\

\noindent{\Large\bfseries Appendix}
\typeout{Appendix}

\noindent 
In order to proof that an inhomogeneous $z(t,r)$ is possible in
the \KS\ model with dust one has to discuss the equations of
motion of the general spherical symmetric model. Its metric can
be parametrized by
\begin{equation}
  ds^2 = -N^2(t,r)dt^2 + L^2(t,r) dr^2 + R^2(t,r) d\Omega^2
\end{equation}
where a vanishing shift--function has been chosen. One nevertheless 
has to satisfy the supermomentum constraint
\begin{equation}
  0\ =\ -L\left(\frac{R\dot R}{N}\right)' + \frac{R'}{R}(LR)\dot{}
\ .
\end{equation}
Here and in the following all equations are presented with the
fields and their velocities instead of their momenta. According
to this constraint a homogeneous surface measure $R(t,r)\to
R(t)=b(t)$ necessitates a homogeneous lapse function $N(t,r)\to
N(t)$. In the following the gauge fixing $N=R$ similar to the one
in the main text is chosen, but the homogeneity of $N,R$ is not
yet enforced.

The gravitational Hamiltonian reads
\begin{equation}
  0\ =\ \frac{L\dot R^2}{R^2} - 2\frac{\dot R^2}{R^2}(LR)\dot{} +
  2R\left(\frac{R'}{L}\right)' + \frac{R'^2}{L} - L + \Lambda LR^2
\ ,
\end{equation}
and the equation of motion are given by
\begin{equation}
  \ddot{R}\ =\ \frac{\dot R^2}{2R}
  - \frac{R}{2} + \frac{3RR'^2}{2L^2} + \frac{\Lambda}{2} R^3
\ ,
\end{equation}
\begin{equation}
  (LR)\ddot{}\ =\ \frac{\dot R}{R}(LR)\dot{} -
  \frac{L\dot R^2}{R} + R^2\left(\frac{R'}{L}\right)' +
   \Lambda LR^2 + R\left(\frac{RR'}{L}\right)'
\ .
\end{equation}
The homogeneity requirement $R(t,r)\to R(t)=b(t)$ obviously leads
to an equation of motion for $b(t)$ which is independent of any
other variable. One can thus insert a solution of it into one of
the other equations in order to determine $L(t,r)$. Furthermore,
since $L'$ does only appear in combination with $R'$ all partial
derivatives cancel and the equations reduce to that of the
inhomogeneous \KS\ spacetime.

A non--vanishing pressure of a matter source would lead to an
interaction of neighboring points. Thus, only incoherent or
pressureless dust might serve as a matter source in this model.
Since according to the Bianchi identities the dust flow is
geodesic one may consider radial moving dust. Furthermore, by the
other part of the Bianchi identities one obtains matter
conservation $\rho(t,r)L(t,r)R^2(t,r)=c_{\mathrm m}(r)$. This
general case is known as the Tolman model and in the case
$R(t,r)\to b(t)$ one gets the inhomogeneous \KS\ model. In the
above equations there will be an additional term in the
Hamiltonian (the potential is supplemented by $- c_{\mathrm
  m}(r)$) but the equations of motion remain unchanged by the
dust. Obviously, there are no further complications to the
inhomogeneity argument due to dust.



\begin{thebibliography}{10}

\bibitem{zeh_book}
H.D.~Zeh,
\newblock {\bf The Physical Basis of the Direction of Time},
\newblock Springer, Heidelberg 1992,
\newblock 2nd ed.

\bibitem{barbour_94a}
J.~B. Barbour,
\newblock {\em The Emergence of Time and Its Arrow form Timelessness},
\newblock In {\bf Physical Origins of Time Asymmetry}, edited by 
  J.J.~Halliwell, J.~Perez-Mercader, and W.H.~Zurek, Cambridge University
  Press, Cambridge 1994.

\bibitem{coleman_88}
S.~Coleman,
\newblock {\em Black Holes as Red Herrings: Topological fluctuations
  and the loss of quantum coherence},
\newblock Nucl. Phys. {\bf B307} (1988) 867.

\bibitem{hawking_90a}
S.W.~Hawking,
\newblock {\em Do Wormholes fix the Constants of Nature},
\newblock Nucl. Phys. {\bf B335} (1990) 155.

\bibitem{fischler_etal_90b}
W.~Fischler, D.~Morgan, and J.~Polchinski,
\newblock {\em Quantization of false vacuum bubbles: A Hamiltonian 
  treatment of gravitational tunneling},
\newblock Phys. Rev. {\bf D42} (1990) 4042.

\bibitem{farhi_etal_90}
E.~Farhi, A.H.~Guth, and J.~Guven,
\newblock {\em Is it possible to create a universe in the laboratory
  by quantum tunneling?},
\newblock Nucl. Phys. {\bf B339} (1990) 417.

\bibitem{dabrowski_larsen_95a}
M.P.~D\c{a}browski and A.L.~Larsen,
\newblock {\em Quantum tunneling effect in oscillating Friedman cosmology},
\newblock Phys.~Rev. {\bf D52} (1995) 3424.

\bibitem{rubakov_84}
V.A.~Rubakov,
\newblock {\em Quantum Mechanics in the Tunneling Universe},
\newblock Phys. Lett. {\bf B148} (1984) 280.

\bibitem{vilenkin_82}
A.~Vilenkin,
\newblock {\em Uniqueness of the tunneling wave function of the Universe},
\newblock Phys. Lett. {\bf B117} (1982) 25.

\bibitem{haroch_kleppner_89}
S.~Haroch and D.~Kleppner,
\newblock {\em Cavity quantum Electrodynamics},
\newblock Physics Today {\bf 42} (1989) 24.

\bibitem{kiefer_zeh_95}
C.~Kiefer and H.D.~Zeh,
\newblock {\em Arrow of time in a recollapsing quantum Universe},
\newblock Phys.~Rev. {\bf D51} (1995) 4145.

\bibitem{caldeira_leggett_83b}
A.O.~Caldeira and A.J.~Leggett,
\newblock {\em Quantum Tunneling in a Dissipative System},
\newblock Ann. Phys.~(N.~Y.) {\bf 149} (1983) 374.

\bibitem{brown_kuchar_95a}
J.D.~Brown and K.V.~Kucha\v{r},
\newblock {\em Dust as a Standard of Space and Time in Canonical Quantum
  Gravity},
\newblock Phys.~Rev. {\bf D51} (1995) 5600.

\bibitem{conradi_95a}
H.-D.~Conradi,
\newblock {\em Kantowski--Sachs like models in quantum cosmology},
\newblock Class.~Quantum~Grav. {\bf 12} (1995) 2423.

\bibitem{laflamme_shellard_87}
R.~Laflamme and E.P.S.~Shellard,
\newblock {\em Quantum cosmology and recollapse},
\newblock Phys.~Rev. {\bf D35} (1987) 2315.

\bibitem{ellis_67}
G.F.R.~Ellis,
\newblock {\em Dynamics of Pressure--Free Matter in General Relativity},
\newblock J. Math. Phys. {\bf 8} (1967) 1171.

\bibitem{conradi_etal_96}
H.-D.~Conradi, A.~Drecker, and H.~Wissowski,
\newblock {\em Notes on Kantowski--Sachs models},
\newblock in preparation.

\bibitem{kiefer_88b}
C.~Kiefer,
\newblock {\em Wave packets in mini-superspace},
\newblock Phys. Rev. {\bf D38} (1988) 1761.

\end{thebibliography}

\end{document}